\documentclass[12pt]{article}
\usepackage{graphicx}
\usepackage{lscape}
\usepackage{citesort}
\usepackage{amssymb}
\usepackage{appendix}
\usepackage{multirow}

\begin{document}
\def\qq{\langle \bar q q \rangle}
\def\uu{\langle \bar u u \rangle}
\def\dd{\langle \bar d d \rangle}
\def\sp{\langle \bar s s \rangle}
\def\GG{\langle g_s^2 G^2 \rangle}
\def\Tr{\mbox{Tr}}

\def\ds{\displaystyle}
\def\beq{\begin{equation}}
\def\eeq{\end{equation}}
\def\bea{\begin{eqnarray}}
\def\eea{\end{eqnarray}}
\def\beeq{\begin{eqnarray}}
\def\eeeq{\end{eqnarray}}
\def\ve{\vert}
\def\vel{\left|}
\def\ver{\right|}
\def\nnb{\nonumber}
\def\ga{\left(}
\def\dr{\right)}
\def\aga{\left\{}
\def\adr{\right\}}
\def\lla{\left<}
\def\rra{\right>}
\def\rar{\rightarrow}
\def\lrar{\leftrightarrow}  
\def\nnb{\nonumber}
\def\la{\langle}
\def\ra{\rangle}
\def\ba{\begin{array}}
\def\ea{\end{array}}
\def\tr{\mbox{Tr}}
\def\ssp{{\Sigma^{*+}}}
\def\sso{{\Sigma^{*0}}}
\def\ssm{{\Sigma^{*-}}}
\def\xis0{{\Xi^{*0}}}
\def\xism{{\Xi^{*-}}}
\def\qs{\la \bar s s \ra}
\def\qu{\la \bar u u \ra}
\def\qd{\la \bar d d \ra}
\def\qq{\la \bar q q \ra}
\def\gGgG{\la g^2 G^2 \ra}
\def\q{\gamma_5 \not\!q}
\def\x{\gamma_5 \not\!x}
\def\g5{\gamma_5}
\def\sb{S_Q^{cf}}
\def\sd{S_d^{be}}
\def\su{S_u^{ad}}
\def\sbp{{S}_Q^{'cf}}
\def\sdp{{S}_d^{'be}}
\def\sup{{S}_u^{'ad}}
\def\ssp{{S}_s^{'??}}

\def\sig{\sigma_{\mu \nu} \gamma_5 p^\mu q^\nu}
\def\fo{f_0(\frac{s_0}{M^2})}
\def\ffi{f_1(\frac{s_0}{M^2})}
\def\fii{f_2(\frac{s_0}{M^2})}
\def\O{{\cal O}}
\def\sl{{\Sigma^0 \Lambda}}
\def\es{\!\!\! &=& \!\!\!}
\def\ap{\!\!\! &\approx& \!\!\!}
\def\ar{&+& \!\!\!}
\def\ek{&-& \!\!\!}
\def\kek{\!\!\!&-& \!\!\!}
\def\cp{&\times& \!\!\!}
\def\se{\!\!\! &\simeq& \!\!\!}
\def\eqv{&\equiv& \!\!\!}
\def\kpm{&\pm& \!\!\!}
\def\kmp{&\mp& \!\!\!}
\def\mcdot{\!\cdot\!}
\def\erar{&\rightarrow&}


\def\simlt{\stackrel{<}{{}_\sim}}
\def\simgt{\stackrel{>}{{}_\sim}}


\renewcommand{\textfraction}{0.2}    
\renewcommand{\topfraction}{0.8}   

\renewcommand{\bottomfraction}{0.4}   
\renewcommand{\floatpagefraction}{0.8}
\newcommand\mysection{\setcounter{equation}{0}\section}

\def\baeq{\begin{appeq}}     \def\eaeq{\end{appeq}}  
\def\baeeq{\begin{appeeq}}   \def\eaeeq{\end{appeeq}}
\newenvironment{appeq}{\beq}{\eeq}   
\newenvironment{appeeq}{\beeq}{\eeeq}
\def\bAPP#1#2{
 \markright{APPENDIX #1}
 \addcontentsline{toc}{section}{Appendix #1: #2}
 \medskip
 \medskip
 \begin{center}      {\bf\LARGE Appendix #1 :}{\quad\Large\bf #2}
\end{center}
 \renewcommand{\thesection}{#1.\arabic{section}}
\setcounter{equation}{0}
        \renewcommand{\thehran}{#1.\arabic{hran}}
\renewenvironment{appeq}
  {  \renewcommand{\theequation}{#1.\arabic{equation}}
     \beq
  }{\eeq}
\renewenvironment{appeeq}
  {  \renewcommand{\theequation}{#1.\arabic{equation}}
     \beeq
  }{\eeeq}
\nopagebreak \noindent}

\def\eAPP{\renewcommand{\thehran}{\thesection.\arabic{hran}}}

\renewcommand{\theequation}{\arabic{equation}}
\newcounter{hran}
\renewcommand{\thehran}{\thesection.\arabic{hran}}

\def\bmini{\setcounter{hran}{\value{equation}}
\refstepcounter{hran}\setcounter{equation}{0}
\renewcommand{\theequation}{\thehran\alph{equation}}\begin{eqnarray}}
\def\bminiG#1{\setcounter{hran}{\value{equation}}
\refstepcounter{hran}\setcounter{equation}{-1}
\renewcommand{\theequation}{\thehran\alph{equation}}
\refstepcounter{equation}\label{#1}\begin{eqnarray}}


\newskip\humongous \humongous=0pt plus 1000pt minus 1000pt
\def\caja{\mathsurround=0pt}


\title{
         {\Large
                 {\bf
Pseudoscalar--meson decuplet--baryon coupling constants 
in light cone QCD 
                 }
         }
      }

\author{\vspace{1cm}\\
{\small T. M. Aliev \thanks
{e-mail: taliev@metu.edu.tr}~\footnote{permanent address:Institute
of Physics,Baku,Azerbaijan}\,\,,
K. Azizi \thanks
{e-mail: e146342@metu.edu.tr}\,\,,
A. \"{O}zpineci \thanks
{e-mail: ozpineci@p409a.physics.metu.edu.tr}\,\,,
M. Savc{\i} \thanks
{e-mail: savci@metu.edu.tr}} \\
{\small Physics Department, Middle East Technical University,
06531 Ankara, Turkey} }

\date{}

\begin{titlepage}
\maketitle
\thispagestyle{empty}

\begin{abstract}
Taking into account the 
$SU(3)_f$ breaking effects, the strong coupling constants of the $\pi$, $K$ and $\eta$ mesons with decuplet
baryons are calculated within light cone QCD sum rules method. It is shown that all coupling 
constants, even in the case of $SU(3)_f$ breaking, are described in terms of only one universal function. It is shown that for  $\Xi^{\ast 0} \rightarrow \Xi^{\ast 0} \eta $  transition violation of $SU(3)_f$ symmetry is very large and for other channels when $SU(3)_f$ symmetry is violated, its maximum value  constitutes $10\%\div15\%$.
\end{abstract}

~~~PACS number(s): 11.55.Hx, 13.75.GX, 13.75.Jz, 14.20.-c, 14.40.Aq
\end{titlepage}

\section{Introduction}

Excited experimental results are obtained on pion and kaon photo-- and
electric--production of nucleon during last several years. These
experiments are performed at different centers, such as MAMI, MIT, Bates,
BNL and Jefferson Laboratories. To study the properties of the resonances
from the existing data, the coupling constants of $\pi$, $K$ and $\eta$
mesons with baryon resonances are needed.

In extracting the properties of baryon resonances, the hadronic reactions 
also play an important role. Therefore, for a more accurate description of
the experimental data, reliable determination of the strong coupling
constants of pseudoscalar mesons is needed. Calculation of the strong
coupling constants of pseudoscalar mesons with baryons (BBP) using the
fundamental theory of strong interactions, QCD, constitutes a very important
problem. The strong coupling constants of BBP belongs to
the nonperturbative sector of QCD and for estimating these couplings, we need
som nonperturbative approachs. Among all nonperturbative approaches, 
the most predictive and powerful one is the QCD sum rules method
\cite{Rde01}. In the present work, we calculate the strong coupling constants
of the pseudoscalar mesons with the decuplet baryons within the framework of
the light cone QCD sum rules (LCSR) method. In this method, the operator
product expansion is performed over twist rather than dimension of the
operators, which is carried out in the traditional sum rules. In the LCSR,
there appears matrix elements of the non--local operators between the vacuum 
and the corresponding one--particle state, which are defined in terms of
the, so--called, distribution amplitudes (DAs). These DAs are the main
nonperturbative parameters of the LCSR method (more about LCSR can be found
in \cite{Rde02,Rde03}). Note that, the coupling constants of pseudoscalar and vector mesons with
octet baryons is investigated within the framework of the LCSR in
\cite{Rde04}, and \cite{Rde05}, respectively.

The paper is organized as follows. In section 2, the the strong coupling
constants
of the pseudoscalar mesons with the decuplet baryons are calculated within 
the framework of the LCSR method, and relations between these coupling
constants are obtained where $SU(3)_f$ symmetry breaking takes place. In
section 3, the numerical analysis of the obtained sum rules for the 
pseudoscalar--meson decuplet--baryon coupling constants is performed.

\section{Light cone QCD sum rules for the pseudo scalar--meson 
decuplet--baryon coupling constants}
In this section, we obtain LCSR for the pseudoscalar--meson decuplet--baryon
coupling constants. For this aim, we consider the following correlation
function
\bea
\label{ede01}
\Pi_{\mu\nu}^{B_1 \rar B_2 {\cal P}} = i \int d^4x e^{ipx} \lla {\cal P}(q) \vel {\cal T} \left\{
\eta_\mu^{B_2} (x) \bar{\eta}_\nu^{B_1} (0) \right\} \ver 0 \rra~,
\eea
where ${\cal P}(q)$ is the pseudoscalar--meson with momentum $q$, $\eta^{B}_\mu$ is the
interpolating current of the considered decuplet baryon. The sum rules for
the above--mentioned correlation function can be obtained, on the one side,
by calculating it in terms of the physical states of hadrons
(phenomenological part), and on the other side, calculating it at $p^2 \rar
- \infty$ in the deep Euclidean region in terms of quarks and gluons
(theoretical part), and equating both representations through the dispersion
relations.

Firstly, let us concentrate on the calculation of the phenomenological side
of the correlation function (\ref{ede01}). The phenomenological part can be
obtained by inserting a complete set of baryon states having the same
quantum numbers as the interpolating current $\eta^B_\mu$. Isolating the ground
state of baryons, we obtain
\bea
\label{ede02}
\Pi_{\mu\nu}^{B_1 \rar B_2 {\cal P}} =  {\lla 0 \vel \eta_\mu^{B_2} \ver
B_2(p_2) \rra \over p_2^2-m_2^2} \lla B_2(p_2) {\cal P}(q) \vel \right.
B_1(p_1) \rra {\lla B_1(p_1) \vel \bar{\eta}_\nu^{B_1} \ver
0 \rra \over p_1^2-m_1^2} + \cdots~,
\eea
where $p_1 = p_2+q$, $m_i$ is the mass of baryon $B_i$, and $\cdots$
represents the contributions of the higher states and the continuum.

The matrix elements of the interpolating current between vacuum and the
hadron states is determined as:
\bea
\label{ede03} 
\lla 0 \vel \eta_\mu \ver B(p,s) \rra = \lambda_B u_\mu (p,s) ~,
\eea
where $\lambda_B$ is the overlap amplitude, and $u_\mu(p,s)$ is the
Rarita--Schwinger tensor spinor with spin $s$. The matrix element $\lla
B_2(p_2) {\cal P}(q) \vel \right. B_1(p_1) \rra$ is parametrized as:

\bea
\label{ede04}
\lla B_2(p_2) {\cal P}(q) \vel \right. B_1(p_1) \rra = g_{B_1 B_2 {\cal P}} \bar{u}_\alpha (p_2) 
\gamma_5 u^\alpha (p_1) ~,
\eea

In order to obtain the expression for the phenomenological part of the
correlation function, the summation over the spins of the Rarita--Schwinger
fields is performed, i.e.,     
\bea
\label{ede05}
\sum_s u_\mu (p,s) \bar{u}_\nu (p,s) =( {\rlap/p +m } )\Bigg(
- g_{\mu\nu} + {1\over 3} \gamma_\mu \gamma_\nu - {2 p_\mu p_\nu \over 3
m^2} - {p_\mu \gamma_\nu - p_\nu \gamma_\mu \over 3 m} \Bigg)~.
\eea

In principle, Eqs.(\ref{ede02}--{\ref{ede05}) allow us to write down
phenomenological part of the correlation function. However, here following two principal
problems appear: 1) not all Lorentz structures are independent; 2) not only
spin--$3/2$, but also spin--$1/2$ states contribute. Indeed, the matrix element of the current $\eta_\mu$, sandwiched between the vacuum 
and the spin--$1/2$ states, is different than zero and determined in the
following way:
\bea
\label{ede06}
\lla 0 \vel \eta_\mu \ver B(p,s=1/2)\rra = A (4 p_\mu - m
\gamma_\mu) u(p,s=1/2)~,
\eea
where the condition $\gamma_\mu \eta^\mu = 0$ has been used.

There are two different alternatives to remove the unwanted spin--$/1/2$ contribution 
and take into account only the independent structures: 1) ordering the Dirac
matrices in a specific way and eliminate the ones that receive
contributions from spin--$/1/2$ states; 2) introduce projection operators
for the spin--$3/2$, that do not contain spin--$1/2$ contribution.

In the present work, we have used the first approach and choose the 
$\gamma_\mu \rlap/p  \rlap/q \gamma_\nu\gamma_5$ ordering of the Dirac
matrices. Having chosen this ordering for the Dirac matrices, we obtain
\bea
\label{ede07}
\Pi_{\mu\nu} \es { \lambda_{B_1} \lambda_{B_2} g_{B_1 B_2 {\cal P}} \over
(p_1^2-m_1^2) (p_2^2 -  m_2^2)} \Big( g_{\mu\nu}  \rlap/p  \rlap/q \gamma_5+
\mbox{\rm other structures with
$\gamma_\mu$ at the beginning and} \nnb \\
&&\mbox{$\gamma_\nu$ at the end, or terms that are proportional to $p_{1\nu}$ or
$p_{2\mu}$} \Big)~.
\eea
 
The advantage of choosing the structure $g_{\mu\nu}  \rlap/p 
\rlap/q\gamma_5$ is in the fact that, the spin--$1/2$ states do not give contribution 
to this structure. This fact immediately follows from Eq. (\ref{ede06}),
which tells that spin--$1/2$ states contribution is proportional to $p_\mu$
or $\gamma_\mu$.

In order to calculate the theoretical part of the correlation function
(\ref{ede01}) from the QCD side, we need the explicit expressions of the
interpolating currents of the decuplet baryons. The interpolating currents
have the following forms \cite{Rde06}:

\bea
\label{ede08}
\eta_\mu = A \epsilon^{abc} \Big[ \Big( q_1^{aT} C \gamma_\mu q_2^b \Big)
q_3^c + \Big( q_2^{aT} C \gamma_\mu q_3^b \Big) q_1^c +
\Big( q_3^{aT} C \gamma_\mu q_1^b \Big) q_2^c \Big] ~,
\eea
where $a,b,c$ are the color indices and $C$ is the charge conjugation
operator. The values of $A$ and the quark flavors $q_1$, $q_2$ and $q_3$ for
each decuplet baryon are presented in Table--1.


\begin{table}[h]

\renewcommand{\arraystretch}{1.3}
\addtolength{\arraycolsep}{-0.5pt}
\small
$$
\begin{array}{|l|c|c|c|c|}
\hline \hline  
 & A & q_1 & q_2 & q_3 \\  \hline
 \Sigma^{\ast 0} & \phantom{-}\sqrt{2/3}  & u & d & s  \\
 \Sigma^{\ast +} &          - \sqrt{1/3}  & u & u & s  \\
 \Sigma^{\ast -} &          - \sqrt{1/3}  & d & d & s  \\
 \Delta^{+}      & \phantom{-}\sqrt{1/3}  & u & u & d  \\
 \Delta^{++}     &        1               & u & u & u  \\
 \Delta^{0}      & \phantom{-}\sqrt{1/3}  & d & d & u  \\
 \Delta^{-}      &        1               & d & d & d  \\
 \Xi^{\ast 0}    & \phantom{-}\sqrt{1/3}  & s & s & u  \\
 \Xi^{\ast -}    & \phantom{-}\sqrt{1/3}  & s & s & d  \\
 \Omega^{-}      &        1               & s & s & s  \\
\hline \hline
\end{array}
$$
\caption{The values of $A$ and the quark flavors $q_1$, $q_2$ and $q_3$} 
\renewcommand{\arraystretch}{1}
\addtolength{\arraycolsep}{-1.0pt}

\end{table}

Before presenting detailed calculation of the correlation function from the
QCD side for determination of the coupling constants of pseudoscalar--mesons
with decuplet--baryons, let us establish the relation among the correlation 
functions, more precisely, relations among the coefficients of the invariant
functions for the structure $g_{\mu\nu}  \rlap/p  \rlap/q\gamma_5$. For this
aim, we will follow the works of \cite{Rde04,Rde08}, and we will show that
all correlation functions which describe the strong coupling constants of
pseudoscalar--mesons with decuplet--baryons can be written in terms of only
one invariant function. It should especially be noted that the approach we
present below automatically takes into account the $SU(3)_f$ symmetry
breaking effects.

In obtaining relations among the invariant functions, similar to works
\cite{Rde04,Rde05}, we start by considering the the correlation function
describing $\Sigma^{\ast 0} \rar \Sigma^{\ast 0}\pi^0$ transition. this
correlation function can formally be written in the following form:
\bea
\label{ede09}
\Pi^{\Sigma^{\ast 0} \rar \Sigma^{\ast 0}\pi^0} = g_{\pi uu} \Pi_1(u,d,s) +
g_{\pi dd} \Pi_1^\prime (u,d,s) + g_{\pi ss} \Pi_2(u,d,s)~,
\eea
where $\pi^0$ current can formally be written as
\bea
\label{ede10}
J = \sum_{q=u,d,s} g_{\pi qq} \, \bar{q} \gamma_5 q~,
\eea
where $g_{\pi^0 uu} = - g_{\pi^0 dd} = 1/\sqrt{2}$ and $g_{\pi^0 ss} = 0$
for $\pi^0$ meson. The functions $\Pi_1$, $\Pi_1^\prime$ and $\Pi_2$
describe radiation $\pi^0$ meson from $u$, $d$ and $s$ quarks of the
$\Sigma^{\ast 0}$ baryon, respectively.

The interpolating current $\eta^{\Sigma^{\ast 0}}$ is symmetric under the
change $u \lrar d$, and therefore $\Pi_1^\prime(u,d,s) = \Pi_1(d,u,s)$.
Hence, Eq. (\ref{ede09}) can be written as:
\bea
\label{ede11}
\Pi^{\Sigma^{\ast 0} \rar \Sigma^{\ast 0}\pi^0} = {1\over \sqrt{2}}
\Big[ \Pi_1(u,d,s) -  \Pi_1(d,u,s) \Big]~.
\eea

For convenience, let us introduce the notations
\bea
\label{ede12}
\Pi_1(u,d,s) \es \lla \bar{u}u \vel \Sigma^{\ast 0} \Sigma^{\ast 0} \ver 0
\rra~, \nnb \\
\Pi_2(u,d,s) \es \lla \bar{s}s \vel \Sigma^{\ast 0} \Sigma^{\ast 0} \ver 0
\rra~.
\eea
Obviously, $\Pi_2 \equiv 0$ for the transition $\Sigma^{\ast 0} \rar 
\Sigma^{\ast 0}\pi^0$. 

In the transition with $\eta$ meson, the situation is
more complicated, since strange quark is in the quark content of the $\eta$
meson. In the present work, we neglect the mixing between the $\eta$ and
$\eta^\prime$ mesons and $\eta$ meson current is taken to have the following
form:
\bea
\label{ede13}
J_\eta = {1\over \sqrt{6}} ( \bar{u} \gamma_5 u + \bar{d} \gamma_5 d
- 2 \bar{s} \gamma_5 s )~.
\eea
A simple analysis shows that the $\Sigma^{\ast 0} \rar \Sigma^{\ast 0}\eta$
transition has the similar form as is given in Eq. (\ref{ede09})
\bea
\label{ede14}
\Pi^{\Sigma^{\ast 0} \rar \Sigma^{\ast 0}\eta} = g_{\eta uu} \Pi_1(u,d,s) +
g_{\eta dd} \Pi_1^\prime (u,d,s) + g_{\eta ss} \Pi_2(u,d,s)~.
\eea
Using the definition given in Eq. (\ref{ede12}), one can easily show that
\bea
\label{ede15}
\Pi_2(u,d,s) = \Pi_1(s,d,u)~. 
\eea
For this reason, using Eqs. (\ref{ede13}) and (\ref{ede15}), we get from Eq.
(\ref{ede14})
\bea
\label{ede16}
\Pi^{\Sigma^{\ast 0} \rar \Sigma^{\ast 0}\eta} = {1\over \sqrt{6}} 
\Big[ \Pi_1(u,d,s) + \Pi_1(d,u,s) - 2 \Pi_1(s,d,u) \Big]~.
\eea

The invariant function describing the $\Sigma^{\ast +} \rar \Sigma^{\ast
+}\pi^0$ transition can be obtained from Eq. (\ref{ede09}) with the help of
the replacements $d \rar u$ in $\Pi_1(u,d,s)$ and using the fact
$\Sigma^{\ast 0} = - \sqrt{2} \Sigma^{\ast +}$, which results in
\bea
\label{ede17}
4 \Pi(u,u,s) = 2 \lla \bar{u}u \vel \Sigma^{\ast +} \Sigma^{\ast +} \ver 0
\rra~.
\eea
The presence of factor 4 on the left--hand side of Eq. (\ref{ede17}) can be
explained as follows. Each $\Sigma^{\ast +}$ contains two $u$ quarks and
therefore there are 4 ways that the $\pi^0$ meson can be radiated. Since
$\Sigma^{\ast +}$ does not contain $d$ quark, for the $\Sigma^{\ast 0} \rar 
\Sigma^{\ast 0}\pi^0$ transition, it can be written from Eq. (\ref{ede09})
that
\bea
\label{ede18}
\Pi^{\Sigma^{\ast +} \rar \Sigma^{\ast +}\pi^0} \es
 g_{\pi^0 uu} \lla \bar{u}u \vel \Sigma^{\ast +} \Sigma^{\ast +} \ver 0 \rra
+ g_{\pi^0 ss} \lla \bar{s}s \vel \Sigma^{\ast +} \Sigma^{\ast +} 
\ver 0 \rra \nnb \\
\es \sqrt{2} \Pi_1(u,u,s)~.
\eea

The result for the $\Sigma^{\ast -} \rar \Sigma^{\ast -}\pi^0$ transition
can easily be obtained by making the replacement $u \rar d$ in Eq.
(\ref{ede09}) and using $\Sigma^{\ast 0} (u\rar d) = \sqrt{2} \Sigma^{\ast
-}$, from which we obtain
\bea
\label{ede19}
\Pi^{\Sigma^{\ast -} \rar \Sigma^{\ast -}\pi^0} \es
 g_{\pi^0 dd} \lla \bar{d}d \vel \Sigma^{\ast -} \Sigma^{\ast -} \ver 0 \rra
+ g_{\pi^0 ss} \lla \bar{s}s \vel \Sigma^{\ast -} \Sigma^{\ast -}
\ver 0 \rra \nnb \\
\es - \sqrt{2} \Pi_1(d,d,s)~.
\eea

In the case of exact isospin symmetry, it follows from Eqs. (\ref{ede11}), 
(\ref{ede18}) and (\ref{ede19}) that
$\Pi^{\Sigma^{\ast 0} \rar \Sigma^{\ast 0}\pi^0} = 0$ and $\Pi^{\Sigma^{\ast +}
\rar \Sigma^{\ast +}\pi^0} = - \Pi^{\Sigma^{\ast -} \rar 
\Sigma^{\ast -}\pi^0}$.

Let us now calculate the invariant function responsible for the
$\Delta^+ \rar \Delta^+ \pi^0$ transition. Since $\Delta^+ = \Sigma^{\ast +}
(s\rar d)$, we get from Eq. (\ref{ede18})
\bea
\label{ede20}
\Pi^{\Delta^+ \rar \Delta^+\pi^0} \es
 g_{\pi^0 uu} \lla \bar{u}u \vel \Sigma^{\ast +} \Sigma^{\ast +} \ver 0 \rra
(s \rar d) 
+ g_{\pi^0 ss} \lla \bar{s}s \vel \Sigma^{\ast +} \Sigma^{\ast +}
\ver 0 \rra (s \rar d) \nnb \\
\es \sqrt{2} \Pi_1(u,u,d) - {1\over \sqrt{2}} \Pi_1(d,u,u)~.
\eea

Similarly, it is not difficult to obtain the relations for the transitions
in which $\Delta^0$, $\Delta^{++}$ and $\Delta^-$ decuplet baryons and
$\pi^0$ meson participate:
\bea
\label{ede21}
\Pi^{\Delta^0 \rar \Delta^0\pi^0} \es \Pi^{\Sigma^{\ast -} \rar 
\Sigma^{\ast -}\pi^0} (s \rar u) \nnb \\
\es- \sqrt{2} \Pi_1(d,d,u) + {1\over \sqrt{2}} 
\Pi_1(u,d,d)~, \nnb \\ \nnb \\
\Pi^{\Delta^{++} \rar \Delta^{++}\pi^0} \es \Pi^{\Sigma^{\ast +} \rar 
\Sigma^{\ast +}\pi^0} (s \rar u) \nnb \\
\es {3\over \sqrt{2}} \Pi_1(u,u,u)~, \nnb \\ \nnb \\
\Pi^{\Delta^- \rar \Delta^-\pi^0} \es \Pi^{\Sigma^{\ast -} \rar 
\Sigma^{\ast -}\pi^0} (s \rar d) \nnb \\        
\es- {3\over \sqrt{2}} \Pi_1(d,d,d)~, \nnb \\ \nnb \\
\Pi^{\Xi^0 \rar \Xi^0\pi^0} \es   {1\over \sqrt{2}} \
\Pi_1(u,s,s)~, \nnb \\ \nnb \\
\Pi^{\Xi^- \rar \Xi^-\pi^0} \es - {1\over \sqrt{2}} \
\Pi_1(d,s,s)~.
\eea

We can proceed now to obtain similar relations in the presence of charged
$\pi$ meson. In order to obtain these relations, we consider the matrix
element $\lla \bar{d}d \vel \Sigma^{\ast 0} \Sigma^{\ast 0} \ver 0 \rra$,
where $d$ quarks from each $\Sigma^{\ast 0}$ form the final $\bar{d}d$ state
and, $u$ and $s$ quarks are the spectators. In the matrix element $\lla
\bar{u}d \vel \Sigma^{\ast +} \Sigma^{\ast 0} \ver 0 \rra$,  $d$ quark from
$\Sigma^{\ast 0}$ and $u$ quark from $\Sigma^{\ast +}$ form the $\bar{u}d$
state and the other $u$ and $s$ quarks are the spectators. For these reasons,
it is natural to expect that these matrix elements should be proportional to
each other. Direct calculations confirm this expectation, i.e.,
\bea
\label{ede22}
\Pi^{\Sigma^{\ast 0} \rar \Sigma^{\ast +} \pi^-} \es \lla \bar{u}d \vel
\Sigma^{\ast +} \Sigma^{\ast 0} \ver 0 \rra =  \sqrt{2} \lla \bar{d}d \vel
\Sigma^{\ast 0} \Sigma^{\ast 0} \ver 0 \rra \nnb \\
\es  \sqrt{2} \Pi_1^\prime (u,d,s) = 
 \sqrt{2} \Pi_1 (d,u,s)~.
\eea

Making the replacement $u \lrar d$ in Eq. (\ref{ede22}), we get
\bea                                       
\label{ede23}
\Pi^{\Sigma^{\ast 0} \rar \Sigma^{\ast -} \pi^+} \es \lla \bar{d}u \vel
\Sigma^{\ast -} \Sigma^{\ast 0} \ver 0 \rra = \sqrt{2} \lla \bar{u}u \vel 
\Sigma^{\ast 0} \Sigma^{\ast 0} \ver 0 \rra \nnb \\
\es \sqrt{2} \Pi_1 (u,d,s)~.
\eea  
Along the same lines of reasoning, similar calculations for for $\Delta$ and
$\Xi$ decuplet baryons are summarized below:
\bea
\label{ede24}
\Pi^{\Xi^{\ast 0} \rar \Xi^{\ast -} \pi^+} \es \lla \bar{d}u \vel
\Xi^{\ast 0} \Xi^{\ast -} \ver 0 \rra = -\sqrt{2}
\lla \bar{u}u \vel \Xi^{\ast 0} \Xi^{\ast 0} \ver 0 \rra
= \Pi_1 (d,s,s)~, \nnb \\
\Pi^{\Xi^{\ast -} \rar \Xi^{\ast 0} \pi^-} \es \lla \bar{u}d \vel
\Xi^{\ast -} \Xi^{\ast 0} \ver 0 \rra 
= \Pi_1 (u,s,s)~, \nnb \\ 
\Pi^{\Delta^+ \rar \Delta^0 \pi^+} \es 2 \Pi_1 (d,d,u)~, \nnb \\
\Pi^{\Delta^{++} \rar \Delta^+ \pi^+} \es  \sqrt{3} \Pi_1 (d,u,u)~, \nnb \\
\Pi^{\Delta^0 \rar \Delta^- \pi^+} \es  \sqrt{3} \Pi_1 (u,d,d)~, \nnb \\
\Pi^{\Delta^0 \rar \Delta^+ \pi^-} \es  2 \Pi_1 (u,u,d)~, \nnb \\
\Pi^{\Delta^+ \rar \Delta^{++} \pi^-} \es  \sqrt{3} \Pi_1 (u,u,u)~, \nnb \\
\Pi^{\Delta^- \rar \Delta^0 \pi^-} \es  \sqrt{3} \Pi_1 (d,d,d)~.
\eea

The correlation function involving $K$ meson can be obtained from the
previous results as follows:

\bea
\label{ede25}
\Pi^{\Xi^{\ast 0} \rar \Sigma^{\ast +} K^-} \es 
\Pi^{\Delta^0 \rar \Delta^+ \pi-} (s \lrar d) = 2 \Pi_1 (u,u,s) \nnb \\
\Pi^{\Xi^{\ast -} \rar \Sigma^{\ast -} K^0} \es 
\Pi^{\Xi^{\ast 0}  \rar \Sigma^+ K^-} (u \rar d) = 2 \Pi_1 (d,d,s) \nnb \\
\Pi^{\Sigma^{\ast +}  \rar \Xi^{\ast 0} K^+} \es
\Pi^{\Xi^{\ast 0}  \rar \Sigma^+ K^-} (u \lrar s) = 2 \Pi_1 (s,s,u)~.
\eea
Remaining correlation functions involving $\pi$ and $K$ mesons are presented
in Appendix--A.
It follows from the results presented above that, all coupling constants of
pseudoscalar mesons with decuplet baryons can be expressed by only one
independent invariant function, which constitutes the main result of the
present work.

Having obtained this result, our next task is the calculation of the
correlation function from QCD side. The correlation function in deep
Euclidean domain $p_1^2\rar - \infty$, $p_2^2 \rar - \infty$, can be
calculated using the operator product expansion. For this purpose the
propagators of light quarks, as well as their distribution amplitudes 
(DA's) are needed. The matrix elements $\lla {\cal P}(q) \vel \bar{q}(x_1) \Gamma
q^\prime (x_2) \ver 0 \rra$ that parametrized in terms of DA's are given in
\cite{Rde09,Rde10,Rde11}:

\bea
\label{ede26}
\lla {\cal P}(p)\vel \bar q(x) \gamma_\mu \gamma_5 q(0)\ver 0 \rra \es 
-i f_{\cal P} q_\mu  \int_0^1 du  e^{i \bar u q x} 
	\left( \varphi_{\cal P}(u) + {1\over 16} m_{\cal P}^2 
x^2 {\Bbb{A}}(u) \right) \nnb \\
\ek {i\over 2} f_{\cal P} m_{\cal P}^2 {x_\mu\over qx} 
\int_0^1 du e^{i \bar u qx} {\Bbb{B}}(u)~,\nnb \\
\lla {\cal P}(p)\vel \bar q(x) i \gamma_5 q(0)\ver 0 \rra \es 
\mu_{\cal P} \int_0^1 du e^{i \bar u qx} \varphi_P(u)~,\nnb \\
\lla {\cal P}(p)\vel \bar q(x) \sigma_{\alpha \beta} \gamma_5 q(0)\ver 0 \rra \es 
{i\over 6} \mu_{\cal P} \left( 1 - \widetilde{\mu}_{\cal P}^2 \right) 
\left( q_\alpha x_\beta - q_\beta x_\alpha\right)
\int_0^1 du e^{i \bar u qx} \varphi_\sigma(u)~,\nnb \\
\lla {\cal P}(p)\vel \bar q(x) \sigma_{\mu \nu} \gamma_5 g_s 
G_{\alpha \beta}(v x) q(0)\ver 0 \rra \es i \mu_{\cal P} \left[
q_\alpha q_\mu \left( g_{\nu \beta} - {1\over qx}(q_\nu x_\beta + 
q_\beta x_\nu) \right) \right. \nnb \\
\ek	q_\alpha q_\nu \left( g_{\mu \beta} - 
{1\over qx}(q_\mu x_\beta + q_\beta x_\mu) \right) \nnb \\
\ek q_\beta q_\mu \left( g_{\nu \alpha} - {1\over qx}
(q_\nu x_\alpha + q_\alpha x_\nu) \right) \nnb \\ 
\ar q_\beta q_\nu \left. \left( g_{\mu \alpha} - 
{1\over qx}(q_\mu x_\alpha + q_\alpha x_\mu) \right) \right] \nnb \\
\cp \int {\cal D} \alpha e^{i (\alpha_{\bar q} + 
v \alpha_g) qx} {\cal T}(\alpha_i)~,\nnb \\
\lla {\cal P}(p)\vel \bar q(x) \gamma_\mu \gamma_5 g_s 
G_{\alpha \beta} (v x) q(0)\ver 0 \rra \es q_\mu (q_\alpha x_\beta - 
q_\beta x_\alpha) {1\over qx} f_{\cal P} m_{\cal P}^2 
\int {\cal D}\alpha e^{i (\alpha_{\bar q} + v \alpha_g) qx} 
{\cal A}_\parallel (\alpha_i) \nnb \\
\ar \left[q_\beta \left( g_{\mu \alpha} - {1\over qx}
(q_\mu x_\alpha + q_\alpha x_\mu) \right) \right. \nnb \\
\ek q_\alpha \left. \left(g_{\mu \beta}  - {1\over qx}
(q_\mu x_\beta + q_\beta x_\mu) \right) \right]
f_{\cal P} m_{\cal P}^2 \nnb \\
\cp \int {\cal D}\alpha e^{i (\alpha_{\bar q} + v \alpha _g) 
q x} {\cal A}_\perp(\alpha_i)~,\nnb \\
\lla {\cal P}(p)\vel \bar q(x) \gamma_\mu i g_s G_{\alpha \beta} 
(v x) q(0)\ver 0 \rra \es q_\mu (q_\alpha x_\beta - q_\beta x_\alpha) 
{1\over qx} f_{\cal P} m_{\cal P}^2 \int {\cal D}\alpha e^{i (\alpha_{\bar q} + 
v \alpha_g) qx} {\cal V}_\parallel (\alpha_i) \nnb \\
\ar \left[q_\beta \left( g_{\mu \alpha} - {1\over qx}
(q_\mu x_\alpha + q_\alpha x_\mu) \right) \right. \nnb \\
\ek q_\alpha \left. \left(g_{\mu \beta}  - {1\over qx}
(q_\mu x_\beta + q_\beta x_\mu) \right) \right] f_{\cal P} m_{\cal P}^2 \nnb \\
	\cp \int {\cal D}\alpha e^{i (\alpha_{\bar q} + 
v \alpha _g) q x} {\cal V}_\perp(\alpha_i)~,
\eea
where 
\bea
\label{nolabel}
\mu_{\cal P} = f_{\cal P} {m_{\cal P}^2\over m_{q_1} + m_{q_2}}~,~~~~~ 
\widetilde{\mu}_{\cal P} = {m_{q_1} + m_{q_2} \over m_{\cal P}}~, \nnb
\eea
and $q_1$ and $q_2$ are the quarks in the meson ${\cal P}$,
${\cal D}\alpha = d\alpha_{\bar q} d\alpha_q d\alpha_g 
\delta(1-\alpha_{\bar q} - \alpha_q - \alpha_g)$, and
and the DA's $\varphi_{\cal P}(u),$ $\Bbb{A}(u),$ $\Bbb{B}(u),$ 
$\varphi_P(u),$ $\varphi_\sigma(u),$ 
${\cal T}(\alpha_i),$ ${\cal A}_\perp(\alpha_i),$ ${\cal A}_\parallel(\alpha_i),$ 
${\cal V}_\perp(\alpha_i)$ and ${\cal V}_\parallel(\alpha_i)$
are functions of definite twist and their expressions are given in 
the next section.

For the calculation of the correlation function, we use the following
expression for the light quark propagator,
\bea
\label{ede27}
S_q(x) \es {i \rlap/x\over 2\pi^2 x^4} - {m_q\over 4 \pi^2 x^2} - 
{\lla \bar q q \rra\over 12} \left(1 - i {m_q\over 4} \rlap/x \right) - 
{x^2\over 192} m_0^2 \lla \bar q q \rra  \left( 1 - 
i {m_q\over 6}\rlap/x \right) \nnb \\ 
&&  - i g_s \int_0^1 du \left[{\rlap/x\over 16 \pi^2 x^2} G_{\mu \nu} (ux) 
\sigma_{\mu \nu} - u x^\mu G_{\mu \nu} (ux) \gamma^\nu 
{i\over 4 \pi^2 x^2} \right. \nnb \\ 
&& \left.
 - i {m_q\over 32 \pi^2} G_{\mu \nu} \sigma^{\mu
 \nu} \left( \ln \left( {-x^2 \Lambda^2\over 4} \right) +
 2 \gamma_E \right) \right]~,
\eea
where $\gamma_E \simeq 0.577$ is the Euler Constant.
In the numerical calculations the scale parameter $\Lambda$ is chosen as
factorization scale, i.e.,  $\Lambda=0.5 \div 1.0~GeV$. This point is
discussed in detail in \cite{Rde12,Rde13}.

Using Eqs. (\ref{ede26}) and (\ref{ede27}) and separating the coefficient
of the structure $g_{\mu\nu}\rlap/p \rlap/q \gamma_5$, the theoretical part of the
correlation function can be calculated straightforwardly. Equating the
coefficients of the structure $g_{\mu\nu}\rlap/p \rlap/q \gamma_5$ from physical and
theoretical parts, and performing Borel transformation in the variables
$p_2^2=p^2$ and $p_1^2=(p+q)^2$ in order to suppress the higher states and
continuum contributions \cite{Rde14,Rde15}, we get the sum rules for the
corresponding pseudoscalar--meson decuplet--baryon coupling constants.

As the result of our calculations, we obtain the following expression for
the invariant function $\Pi_1(u,d,s)$.
 
\bea
\label{ede28}
\Pi_1(u,d,s) \es {1\over 54 \pi^2} M^4 E_1(x) \Big[ 9 f_{\cal P} m_s \phi_{\cal P} (u_0) +
2 (1-\widetilde{\mu}_{\cal P}^2) \mu_{\cal P} \phi_\sigma (u_0) \Big] \nnb \\
\ar {1\over 36 \pi^2} \Big\{f_{\cal P} M^2 E_0(x) \Big[-3 m_{\cal P}^2 m_s \Big( A(u_0) - 4
i(A_\parallel,1-2 v) \Big) - 16 \pi^2 \phi_{\cal P} (u_0) 
( \dd + \sp ) \Big\} \nnb \\
\ar {1\over 216 M^6} \dd \gGgG m_s (-1 + \widetilde{\mu}_{\cal P}^2) \phi_\sigma
(u_0) ( m_0^2 + 2 M^2) \nnb \\
\ar {1\over 7776 \pi^2 M^2} \Big[ 27 f_{\cal P} m_{\cal P}^2 m_s \gGgG \Big( A(u_0) - 4
i(A_\parallel,1-2 v) + 4 i(V_\parallel,1) \Big) \nnb \\
\ar 32 \pi^2 m_0^2 m_s \mu_{\cal P} (1 - \widetilde{\mu}_{\cal P}^2) (\dd + 2 \sp ) 
\phi_\sigma (u_0) \Big] \nnb \\
\ar {1\over 72 \pi^2} \Big[ f_{\cal P} m_s \Big( \gamma_E + \ln {\Lambda^2\over
M^2}\Big) \Big( 24 m_{\cal P}^2 M^2 E_0(x) i(V_\parallel,1) + 
\gGgG \phi_{\cal P} (u_0) \Big) \Big] \nnb \\
\ar {1\over 9} f_{\cal P} m_{\cal P}^2 A(u_0) ( \dd +\sp ) - 4 f_{\cal P} m_{\cal P}^2 (\dd + \sp) \Big[
i(A_\parallel,1-2 v) - i(V_\parallel,1) \Big] \nnb \\
\ar {1\over 324 \pi^2} f_{\cal P} \Big[ - 3 m_s \gGgG + 40 m_0^2 \pi^2 (\dd+\sp)
\phi_{\cal P}(u_0) \nnb \\
\ek {2\over 9} m_s \mu_{\cal P} (1 - \widetilde{\mu}_{\cal P}^2) \dd \phi_\sigma~,
\eea
where
\bea
\label{nolabel}
\mu_{\cal P} \es {f_{\cal P} 
m_{\cal P}^2 \over m_{q_1}+m_{q_2}}~,~~~~~  
\widetilde{\mu}_{\cal P} = {m_{q_1}+m_{q_2} \over m_{\cal P}}~, \nnb   
\eea
and the function $i(\varphi,f(v))$ is defined as follows
\bea
\label{nolabel}
i(\varphi,f(v)) \es \int {\cal D}\alpha_i \int_0^1 dv\,
\varphi(\alpha_{\bar{q}},\alpha_q,\alpha_g) 
f(v) \delta(k-u_0)~, \nnb
\eea
where
\bea
\label{nolabel}
k = \alpha_q + \alpha_g \bar{v}~,~~~~~
u_0 = {M_1^2 \over M_1^2 + M_2^2}~,~~~~~ M^2 = {M_1^2 M_2^2 \over M_1^2 +
M_2^2}~. \nnb
\eea

In calculating the coupling constants of pseudoscalar--mesons with decuplet
baryons, the value of the overlap amplitude $\lambda_B$ of the hadron is
needed. This overlap amplitude is determined from the analysis of the
two--point function which is calculated in \cite{Rde14,Rde15}. Our earlier
considerations reveal that the interpolating currents of decuplet baryons
can all be obtained from the $\Sigma^{\ast 0}$ current, and for this reason
we shall present the result only for the overlap amplitude of 
$\Sigma^{\ast 0}$:

\bea
\label{ede29}
M_{\Sigma^{\ast 0}} \lambda_{\Sigma^{\ast 0}}^2 e^{-{m_\Sigma^2\over M^2}} \es 
\left( \uu + \dd + \sp \right) {M^4\over 9 \pi^2} E_1(x)
- \left( m_u + m_d + m_s\right) {M^6\over 32 \pi^4} E_2(x) \nnb \\ 
\ek \left( \uu + \dd + \sp \right) m_0^2 {M^2\over 18 \pi^2} E_0(x) \nnb \\ 
\ek {2\over 3}\left(1 + {5 m_0^2\over 72 M^2} \right) \left( m_u \dd \sp +
m_d \sp \uu + m_s \dd \uu \right) \nnb \\ 
\ar \left( m_s \dd \sp + m_u \dd \uu + m_d \sp \uu \right) {m_0^2\over 12
M^2}~,
\eea
where $x = s_0/M^2$.

The contribution of the higher states and continuum in $\Pi_1$ are 
subtracted by taking
into account  the following replacements  

\bea
\label{ede30}
e^{-m_{\cal P}^2/4 M^2} M^2 \left( \ln{M^2 \over \Lambda^2} - \gamma_E \right)  
\erar \int_{m_{\cal P}^2/4}^{s_0}  ds  e^{- s/M^2} 
\ln {s-m_{\cal P}^2/4\over \Lambda^2}\nnb \\
e^{-m_{\cal P}^2/4 M^2} \left( \ln{M^2 \over \Lambda^2} - \gamma_E \right) 
\erar \ln {s_0 - m_{\cal P}^2/4\over \Lambda^2} e^{-s_0/M^2} + 
{1\over M^2} \int_{m_{\cal P}^2/4}^{s_0} ds e^{- s/M^2} 
\ln {s-m_{\cal P}^2/4\over \Lambda^2}\nnb \\
e^{-m_{\cal P}^2/4 M^2} {1\over M^2} \left( \ln{M^2 \over \Lambda^2} - 
\gamma_E \right) 
\erar {1\over M^2} \ln 
{s_0 - m_{\cal P}^2/4\over \Lambda^2} e^{-s_0/M^2}
+ {1\over s_0-m_{\cal P}^2/4}  e^{-s_0/M^2} \nnb \\ 
\ar {1\over M^4} \int_{m_{\cal P}^2/4}^{s_0} ds e^{- s/M^2} 
\ln {s-m_{\cal P}^2/4\over \Lambda^2}\nnb \\
e^{-m_{\cal P}^2 / 4 M^2} M^{2n}
\erar {1\over \Gamma(n)} 
\int_{m_{\cal P}^2/4}^{s_0} ds  e^{- s/M^2} \left( s - m_{\cal P}^2/4
\right)^{n-1}~.
\eea 
   
\section{Numerical analysis}

In this section, we present the numerical calculations for the sum rules for
the couplings of the pseudoscalar--mesons with decuplet--baryons. The main
nonperturbative parameters of LCSR are the DA's of the pseudoscalar mesons,
whose explicit forms entering Eq. (\ref{ede26}) are given in
\cite{Rde09,Rde10,Rde11}:
  
\bea
\label{ede31}
\phi_{\cal P}(u) \es 6 u \bar u \left[ 1 + a_1^{\cal P} C_1(2 u -1) + 
a_2^{\cal P} C_2^{3/2}(2 u - 1) \right]~,  \nnb \\
{\cal T}(\alpha_i) \es 360 \eta_3 \alpha_{\bar q} \alpha_q 
\alpha_g^2 \left[ 1 + w_3 {1\over 2} (7 \alpha_g-3) \right]~, \nnb \\
\phi_P(u) \es 1 + \left[ 30 \eta_3 - {5\over 2} 
{1\over \mu_{\cal P}^2}\right] C_2^{1/2}(2 u - 1)~,  \nnb \\ 
\ar	\left( -3 \eta_3 w_3  - {27\over 20} {1\over \mu_{\cal P}^2} - 
{81\over 10} {1\over \mu_{\cal P}^2} a_2^{\cal P} \right) 
C_4^{1/2}(2u-1)~, \nnb \\
\phi_\sigma(u) \es 6 u \bar u \left[ 1 + \left(5 \eta_3 - {1\over 2} \eta_3 w_3 - 
{7\over 20}  \mu_{\cal P}^2 - {3\over 5} \mu_{\cal P}^2 a_2^{\cal P} \right)
C_2^{3/2}(2u-1) \right] ~, \nnb \\
{\cal V}_\parallel(\alpha_i) \es 120 \alpha_q \alpha_{\bar q} \alpha_g 
\left( v_{00} + v_{10} (3 \alpha_g -1) \right) ~, \nnb \\
{\cal A}_\parallel(\alpha_i) \es 120 \alpha_q \alpha_{\bar q} \alpha_g 
\left( 0 + a_{10} (\alpha_q - \alpha_{\bar q}) \right) ~, \nnb \\
{\cal V}_\perp (\alpha_i) \es - 30 \alpha_g^2\left[ h_{00}(1-\alpha_g) + 
h_{01} (\alpha_g(1-\alpha_g)- 6 \alpha_q \alpha_{\bar q}) +
h_{10}(\alpha_g(1-\alpha_g) - {3\over 2} (\alpha_{\bar q}^2+ 
\alpha_q^2)) \right] ~, \nnb \\
{\cal A}_\perp (\alpha_i) \es 30 \alpha_g^2(\alpha_{\bar q} - \alpha_q) 
\left[ h_{00} + h_{01} \alpha_g + {1\over 2} h_{10}(5 \alpha_g-3) \right] ~, \nnb \\
B(u)\es g_{\cal P}(u) - \phi_{\cal P}(u) ~, \nnb \\
g_{\cal P}(u) \es g_0 C_0^{1/2}(2 u - 1) + g_2 C_2^{1/2}(2 u - 1) + 
g_4 C_4^{1/2}(2 u - 1) ~, \nnb \\
\Bbb{A}(u) \es 6 u \bar u \left[{16\over 15} + {24\over 35} a_2^{\cal P}+ 
20 \eta_3 + {20\over 9} \eta_4 +
\left( - {1\over 15}+ {1\over 16}- {7\over 27}\eta_3 w_3 - 
{10\over 27} \eta_4 \right) C_2^{3/2}(2 u - 1)  \right. \nnb \\ 
	\ar \left. \left( - {11\over 210}a_2^{\cal P} - {4\over 135} 
\eta_3w_3 \right)C_4^{3/2}(2 u - 1)\right] ~, \nnb \\
\ar \left( -{18\over 5} a_2^{\cal P} + 21 \eta_4 w_4 \right)
\left[ 2 u^3 (10 - 15 u + 6 u^2) \ln u  \right. \nnb \\
\ar \left. 2 \bar u^3 (10 - 15 \bar u + 6 \bar u ^2) \ln\bar u + 
u \bar u (2 + 13 u \bar u) \right]~,
\eea
where $C_n^k(x)$ are the Gegenbauer polynomials, and   
\bea
\label{ede32}
h_{00}\es v_{00} = - {1\over 3}\eta_4 ~, \nnb \\
a_{10} \es {21\over 8} \eta_4 w_4 - {9\over 20} a_2^{\cal P} ~, \nnb \\
v_{10} \es {21\over 8} \eta_4 w_4 ~, \nnb \\
h_{01} \es {7\over 4}  \eta_4 w_4  - {3\over 20} a_2^{\cal P} ~, \nnb \\
h_{10} \es {7\over 4} \eta_4 w_4 + {3\over 20} a_2^{\cal P} ~, \nnb \\
g_0 \es 1 ~, \nnb \\
g_2 \es 1 + {18\over 7} a_2^{\cal P} + 60 \eta_3  + {20\over 3} \eta_4 ~, \nnb \\
g_4 \es  - {9\over 28} a_2^{\cal P} - 6 \eta_3 w_3~.
\eea
The values of the parameters $a_1^{\cal P}$, $a_2^{\cal P}$,
$\eta_3$, $\eta_4$, $w_3$, and $w_4$ entering Eqs. (\ref{ede32}) are given in 
Table (\ref{param}) for the $\pi$, $K$ and $\eta$ mesons.

\begin{table}
\begin{center}
\begin{tabular}{|c|c|c|c|}
\hline\hline
		&	$\pi$	&	$K$      &  $\eta$	\\
\hline 
$a_1^{\cal P}$	&	$0$	&	$0.050$	 &   0      \\
\hline
$a_2^{\cal P}$	&	$0.44$ 	&	$0.16$	 &  0.2       \\
\hline
$\eta_3$	&	$0.015$	&	$0.015$	 &  0.013       \\
\hline
$\eta_4$	&	$10$	&	$0.6$	 &  0.5         \\
\hline
$w_3$		&	$-3$	&	$-3$	 &  -3          \\
\hline
$w_4$		&	$0.2$	&	$0.2$	 & 0.2         \\
\hline \hline
\end{tabular}
\end{center}
\caption{Parameters of the wave function calculated at the renormalization scale $\mu = 1 ~GeV$}
\label{param}
\end{table}

In the numerical calculations, we set $M_1^2 = M_2^2 = 2 M^2$ due to the fact
that the masses of the initial and final
baryons are close to each other. With this choice, we have   
$u_0 = 1/2$. The values of the other input
parameters entering  the sum rules are: $\lla \bar{q} q \rra = -
(0.24\pm 0.01~GeV)^3$, 
$m_0^2=(0.8 \pm 0.2)~GeV^2$ \cite{Rde14}, $f_\pi=0.131~GeV$, $f_K=0.16~GeV$,
and $f_\eta=0.13~GeV$ \cite{Rde09}.

The sun rules for the coupling constant of pseudoscalar--mesons with
decuplet--baryons contain two auxiliary, namely, Borel parameter $M^2$ and
the continuum threshold $s_0$. Obviously, we need to find such regions of
these parameters where coupling constants are practically independent of
them.

The upper limit of $M^2$ can be found by requiring that the higher states
and continuum contributions to the correlation function should be less than
40--50\% of the total value of the correlation function. The lower bound of
$M^2$ can be obtained by demanding that the contribution of the highest term
with power $1/M^2$ is less than, say, 20--25\% of the highest power of $M^{2}$. 
Using these two conditions, one can find regions of $M^2$ where the results
for the coupling constants are insensitive to the variation of $M^2$. The
value of the continuum threshold is varied between $2.5~GeV^2 \le s_0 \le
4~GeV^2$.

As an example, in Fig. (1), we depict the dependence of the 
$g^{\Sigma^{\ast +} \rar \Sigma^{\ast +} \pi^0}$ coupling constant on $M^2$
at fixed values of the continuum threshold. From this figure, one can see
that the $g^{\Sigma^{\ast +} \rar \Sigma^{\ast +} \pi^0}$ coupling constant
demonstrates good stability to the variation in $M^2$. The numerical results
for the coupling constants of pseudoscalar--mesons with decuplet--baryons
are presented in Table (\ref{results}). Note that in this Table, we give only those
results which are not obtained from each other by $SU(2)$ and isotopic spin
relations. It should be remembered that the sum rules cannot fix the signs of
the residues and for this reason the signs of the couplings are not fixed.
However, they can be fixed if we use $SU(3)_f$ symmetry (more about this
issue, see \cite{Rde04}). 
\begin{figure}[h]
\begin{center}
\caption{The  dependence of the 
$g^{\Sigma^{\ast +} \rar \Sigma^{\ast +} \pi^0}$ coupling constant on $M^2$
at fixed values of the continuum threshold.} \label{fig1}
\includegraphics[width=18cm]{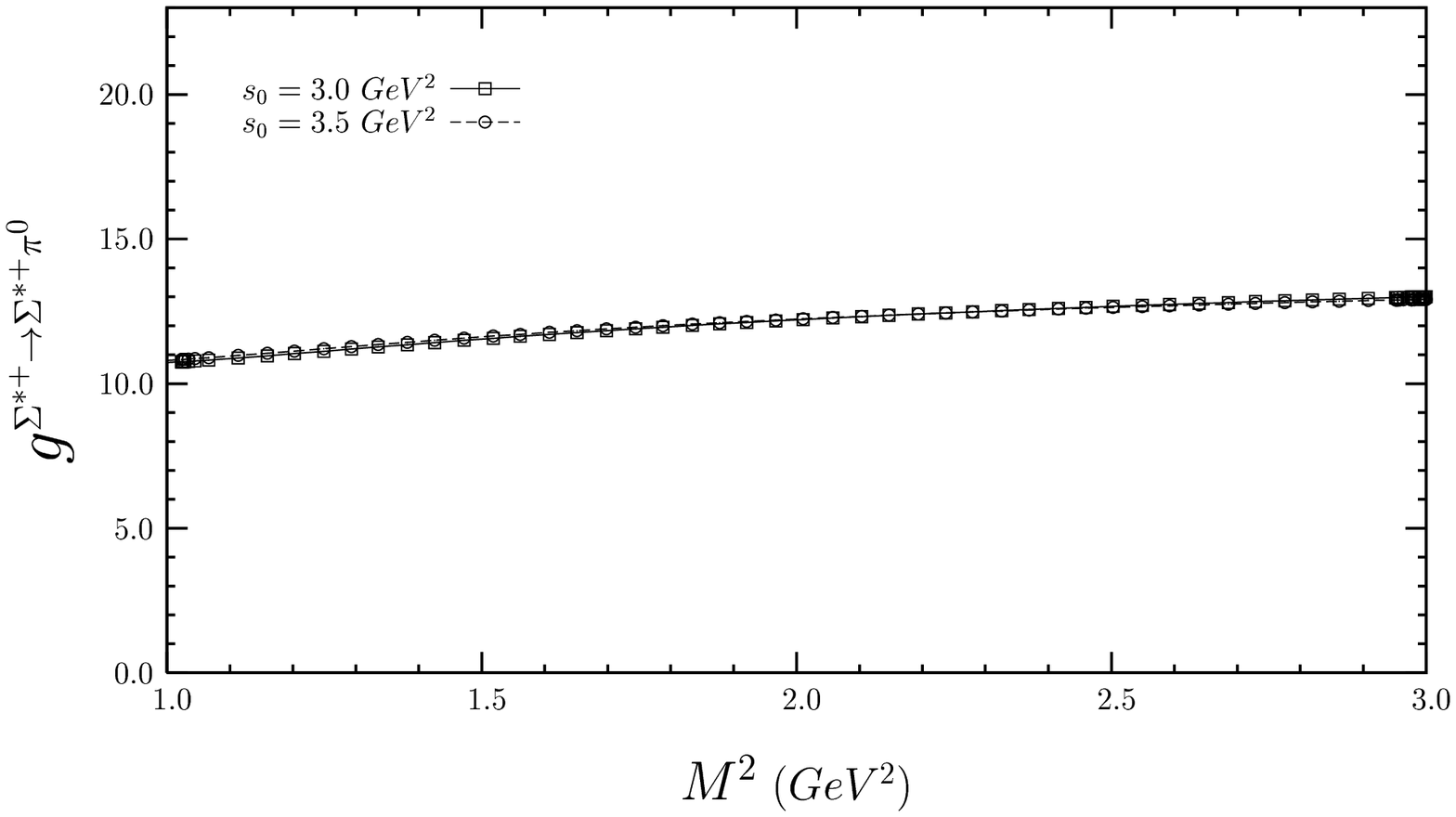}
\end{center}
\end{figure}
From this Table, we can deduce the following conclusions:
\begin{itemize}
 \item In all considered couplings except $\Sigma^{\ast 0} \rightarrow \Sigma^{\ast 0} \eta $ our predictions consist with the $SU(3)_f$ symmetry. Maximum violation of $SU(3)_f$ symmetry is about 15\%.
\item In $SU(3)_f$ symmetry limit coupling constant for $\Sigma^{\ast 0} \rightarrow \Sigma^{\ast 0} \eta $ transition is equal to zero, but our prediction on this constant differs from zero considrably when violation of $SU(3)_f$ symmetry is taken into account. Only for this channel violation of $SU(3)_f$ symmetry is huge. In principle, investigation of this coupling constant can shed light on the structure of $\eta$ meson.
\item Sign of coupling constant of decuplet baryons to K meson and also  $\Xi^{\ast 0} \rightarrow \Xi^{\ast 0} \eta $ is negative, but for all other cases is positive.
\end{itemize}
In summary, cosidering the $SU(3)_f$ symmetry breraking effects, the coupling constants of the decuplet baryons with pseudoscalar  $\pi$, $K$ and $\eta$ mesons have been calculated in the framework of light cone QCD sum rules. It was shown that all aforementiond coupling constants is described with the help of one universal function. We obtained that for $\Xi^{\ast 0} \rightarrow \Xi^{\ast 0} \eta $ transition, violation of  $SU(3)_f$ is very large.

\begin{table}[h]

\renewcommand{\arraystretch}{1.3}
\addtolength{\arraycolsep}{-0.5pt}
\small
$$
\begin{array}{|r@{\rar}l|r@{\pm}l|r@{\pm}l|} 
\hline \hline  
\multicolumn{2}{|c}{\mbox{\rm channel}} &  \multicolumn{2}{c|}{\mbox{\rm Coupling}} &\multicolumn{2}{c|}{\mbox{\rm Coupling in SU(3) limit}} \\ \hline
\Sigma^{\ast +} & \Sigma^{\ast +} \pi^0 &  11.3&1.2  &  $~~~~~~~~~~$11.0&1.2\\
\Delta^+        & \Delta^+ \pi^0        &   5.5&0.8  &  5.5&0.8\\
\Xi^{\ast 0}    & \Xi^{\ast 0} \pi^0    &   5.2&0.3  &  5.5&0.5\\
\Sigma^{\ast 0} & \Delta^+ K^-          & -17.4&1.6  &  -18.0&1.8\\
\Xi^{\ast 0}    & \Sigma^{\ast +} K^-   & -25.4&2.5  &  -27.0&2.8\\
\Sigma^{\ast +} & \Delta^{++} K^-       & -21.2&1.8  &  -22.0&2.0\\
\Omega^-        & \Xi^{\ast 0} K^-      & -20.7&1.6  &  -22.2&2.0\\
\Sigma^{\ast +} & \Xi^{\ast 0} K^+      & -22.2&1.6  &  -27.0&2.8\\
\Sigma^{\ast 0} & \Sigma^{\ast 0} \eta  &  0.65&0.05 & 0.0&0.0\\
\Delta^+        & \Delta^+ \eta         &  12.5&1.5  &  12.6&1.5 \\
\Xi^{\ast 0}    & \Xi^{\ast 0} \eta     & -11.2&1.0  &   -13.2&1.8\\
\hline \hline

\end{array}
$$
\caption{Coupling constants of pseudoscalar--mesons with decuplet--baryons} 
\renewcommand{\arraystretch}{1}
\addtolength{\arraycolsep}{-1.0pt}
\label{results}
\end{table}

\section{Acknowledgment}
 K. A. thanks TUBITAK, Turkish Scientific and Technological
Research Council, for their financial support provided under the
project 108T502.
\newpage

\bAPP{A}{}

In this appendix we present the correlation functions involving $\pi$, $K$ and
$\eta$ mesons which is not given in the main text.

\begin{itemize}
\item Correlation functions for the couplings involving $\pi^+$ meson
\end{itemize}
\baeeq
\Pi^{\Sigma^{\ast +} \rar \Sigma^{\ast 0} \pi^+} \es 
 \sqrt{2} \Pi_1(d,u,s)~. \nnb
\eaeeq
\begin{itemize}
\item Correlation functions for the couplings involving $\pi^-$ meson 
\end{itemize}
\baeeq
\Pi^{\Sigma^{\ast -} \rar \Sigma^{\ast 0} \pi^-} \es
\sqrt{2} \Pi_1(u,d,s)~, \nnb \\
\Pi^{\Delta^0 \rar \Delta^+ \pi^-} \es
2 \Pi_1(u,u,d)~. \nnb
\eaeeq
\begin{itemize}
\item Correlation functions for the couplings involving $K$ meson
\end{itemize}
\baeeq
\Pi^{\Sigma^{\ast 0} \rar \Delta^+ K^-} \es
 \sqrt{2} \Pi_1(s,u,d)~, \nnb \\
\Pi^{\Xi^{\ast -} \rar \Sigma^{\ast 0} K^-} \es
 \sqrt{2} \Pi_1(u,d,s)~, \nnb \\
\Pi^{\Delta^- \rar \Sigma^{\ast 0} K-} \es
0~, \nnb \\
\Pi^{\Sigma^{\ast +} \rar \Delta^{++} K^-} \es
 \sqrt{3} \Pi_1(u,u,u)~, \nnb \\
\Pi^{\Sigma^{\ast -} \rar \Delta^0 K^-} \es
 \Pi_1(s,d,d)~, \nnb \\
\Pi^{\Sigma^{\ast -} \rar \Xi^{\ast 0} K^-} \es
0~, \nnb \\
\Pi^{\Omega^- \rar \Xi^{\ast 0} K^-} \es
 \sqrt{3} \Pi_1(s,s,s)~, \nnb \\
\Pi^{\Delta^0 \rar \Sigma^{\ast +} K^-} \es
0~, \nnb \\
\Pi^{\Delta^+ \rar \Sigma^{\ast 0} K^+} \es
 \sqrt{2} \Pi_1(s,u,d)~, \nnb \\
\Pi^{\Delta^0 \rar \Sigma^{\ast -} K^+} \es
 \Pi_1(s,d,d)~, \nnb \\
\Pi^{\Sigma^{\ast +} \rar \Delta^0 K^+} \es
0~, \nnb \\
\Pi^{\Delta^{++} \rar \Sigma^{\ast +} K^+} \es
 \sqrt{3} \Pi_1(u,u,u)~, \nnb \\
\Pi^{\Xi^{\ast 0} \rar \Sigma^{\ast -} K^+} \es
0~, \nnb \\
\Pi^{\Xi^{\ast 0} \rar \Sigma^{\ast 0} \bar{K}^0} \es
\Pi^{\Sigma^{\ast 0} \rar \Xi^{\ast 0} \bar{K}^0} =
\Pi^{\Xi^{\ast 0} \rar \Sigma^{\ast 0} K^0} =
\Pi^{\Sigma^{\ast 0} \rar \Xi^{\ast 0} K^0} =
\sqrt{2} \Pi_1(d,u,s)~, \nnb \\
\Pi^{\Sigma^{\ast -} \rar \Xi^{\ast -} \bar{K}^0} \es
\Pi^{\Xi^{\ast -} \rar \Sigma^{\ast -} K^0} =
 2 \Pi_1(s,s,d)~, \nnb \\
\Pi^{\Omega^- \rar \Xi^{\ast -} \bar{K}^0} \es
\Pi^{\Xi^{\ast -} \rar \Omega^- \bar{K}^0} =
\Pi^{\Xi^{\ast 0} \rar \Omega^- K^+} = 
\Pi^{\Omega^- \rar \Xi^{\ast -} K^0} = 
\Pi^{\Xi^{\ast -} \rar \Omega^- K^0} =
 \sqrt{3} \Pi_1(s,s,s)~, \nnb \\
\Pi^{\Sigma^{\ast 0} \rar \Delta^0 \bar{K}^0} \es
\Pi^{\Delta^0 \rar \Sigma^{\ast 0} \bar{K}^0} =
\Pi^{\Sigma^{\ast 0} \rar \Delta^0 K^0} =
 \sqrt{2} \Pi_1(s,d,u)~, \nnb \\
\Pi^{\Sigma^{\ast +} \rar \Delta^+ \bar{K}^0} \es
\Pi^{\Delta^+ \rar \Sigma^{\ast +} \bar{K}^0} =
 \Pi_1(s,u,u)~, \nnb \\
\Pi^{\Delta^- \rar \Sigma^{\ast -} \bar{K}^0} \es
\Pi^{\Sigma^{\ast -} \rar \Delta^- \bar{K}^0} =
 \sqrt{3} \Pi_1(s,d,d)~, \nnb \\
\Pi^{\Delta^+ \rar \Sigma^{\ast +} K^0} \es
\Pi^{\Sigma^{\ast +} \rar \Delta^+ K^0} =
 \Pi_1(s,u,u)~, \nnb \\
\Pi^{\Delta^- \rar \Sigma^{\ast -} K^0} \es
\Pi^{\Sigma^{\ast -} \rar \Delta^- K^0} =
 \sqrt{3} \Pi_1(s,d,d)~, \nnb \\
\Pi^{\Delta^0 \rar \Sigma^{\ast 0} K^0} \es
 \sqrt{2} \Pi_1(s,u,d)~, \nnb \\
\Pi^{\Sigma^{\ast -} \rar \Xi^- K^0} \es
 2 \Pi_1(d,d,s)~. \nnb
\eaeeq
\begin{itemize}
\item Correlation functions for the couplings involving $\eta$ meson
\end{itemize}
\baeeq
\Pi^{\Sigma^{\ast 0} \rar \Sigma^{\ast 0} \eta} \es
 {1\over \sqrt{6}} \Big[ \Pi_1(u,d,s) + \Pi_1(d,u,s) - 
2 \Pi_1(s,d,u) \Big]~, \nnb \\
\Pi^{\Sigma^{\ast +} \rar \Sigma^{\ast +} \eta} \es
 {2\over \sqrt{6}} \Big[ \Pi_1(u,u,s) -
 \Pi_1(s,u,u) \Big]~, \nnb \\
\Pi^{\Sigma^{\ast -} \rar \Sigma^{\ast -} \eta} \es
 {2\over \sqrt{6}} \Big[ \Pi_1(d,d,s) -
 \Pi_1(s,d,d) \Big]~, \nnb \\
\Pi^{\Delta^+ \rar \Delta^+ \eta} \es
 {1\over \sqrt{6}} \Big[ 2 \Pi_1(u,u,d) +
 \Pi_1(d,u,u) \Big]~, \nnb \\
\Pi^{\Delta^{++} \rar \Delta^{++} \eta} \es
 {\sqrt{6}\over 2} \Pi_1(u,u,u)~, \nnb \\
\Pi^{\Delta^- \rar \Delta^- \eta} \es           
 {\sqrt{6}\over 2} \Pi_1(d,d,d)~, \nnb \\
\Pi^{\Delta^0 \rar \Delta^0 \eta} \es           
 {1\over \sqrt{6}} \Big[ 2 \Pi_1(d,d,u) +
 \Pi_1(u,d,d) \Big]~, \nnb \\
\Pi^{\Xi^{\ast 0} \rar \Xi^{\ast 0} \eta} \es           
 {1\over \sqrt{6}} \Big[ \Pi_1(u,s,s) -
4 \Pi_1(s,s,u) \Big]~, \nnb \\
\Pi^{\Xi^{\ast -} \rar \Xi^{\ast -} \eta} \es           
 {1\over \sqrt{6}} \Big[ \Pi_1(d,s,s) -
4 \Pi_1(s,s,d) \Big]~. \nnb
\eaeeq

\eAPP

\end{document}